\pdfoutput=1

\documentclass[%
 reprint,
 floatfix,
 widetext,
 amsmath,amssymb,
 aps,
]{revtex4-2}

\usepackage[strict]{changepage}
\usepackage[title]{appendix}
\usepackage{xcolor}
\usepackage{graphicx}
\usepackage{dcolumn}
\usepackage{bm}

\usepackage[utf8]{inputenc}
\usepackage{physics}
\usepackage{amsfonts}
\usepackage{amsmath}
\usepackage{xcolor}

\begin{document}

\preprint{APS/123-QED}

\title{Effect of electronic correlation on topological end-states in finite-size graphene nanoribbons}

\author{Antoine Honet}
\affiliation{%
Department of Physics and Namur Institute of Structured Materials, University of Namur, Rue de Bruxelles 51, 5000 Namur, Belgium
}%

\author{Luc Henrard}
\affiliation{%
Department of Physics and Namur Institute of Structured Materials, University of Namur, Rue de Bruxelles 51, 5000 Namur, Belgium
}%

\author{Vincent Meunier}%
\affiliation{%
Department of Engineering Science and Mechanics, The Pennsylvania State University, University Park, PA, USA
}%

\date{\today}

\begin{abstract}
Finite size armchair graphene nanoribbons (GNR) of different families are theoretically studied using the Hubbard model in both mean-field and GW approximations, including spin correlation effects. It is shown that correlation primarily affect the properties of topological end states of the nanoribbons. A representative structure of each of the three GNR families is considered but the 7-atom width nanoribbon is studied in detail and compared to experimental results, showing a clear improvement when correlations are included. Using on numerically computed local density of states, spin-polarized measurements in scanning tunneling microscopy are also suggested to help distinguish and highlight correlation effects.
 
\begin{description}
\item[Keywords]
Graphene nanorribons, Hubbard model, mean-field approximation, GW approximation, Green's function theory, topological end states
\end{description}
\end{abstract}

\maketitle


\author{Antoine Honet, Luc Henrard and Vincent Meunier}

\section{Introduction}

Since its experimental isolation~\cite{novoselov_electric_2004}, graphene has been extensively studied owing in part to its unique electronic properties~\cite{castro_neto_electronic_2009}. Macroscopically large graphene sheets are semi-metallic, \textit{i.e.} with a zero electronic band gap. However, fragments of graphene of different sizes and shapes can display significantly different electronic properties compared to graphene, for example with the appearance of an electronic band gap. The spatial confinement in graphene fragments can also induce the emergence of magnetic properties~\cite{yazyev_emergence_2010}. In particular, graphene nanoribbons (GNRs) are of special interest since they can be synthesized with atomic precision~\cite{ruffieux_electronic_2012, kimouche_ultra-narrow_2015, sode_electronic_2015, talirz_-surface_2017, talirz_termini_2013, zhang_-surface_2015, lawrence_probing_2020, wang_giant_2016}. The development of synthesis processes and theoretical studies~\cite{son_energy_2006, brey_electronic_2006, lu_competition_2016, hagymasi_entanglement_2016} have opened new research directions where the electronic properties of GNRs can be tuned and engineered by structuring the GNRs. For instance, heterojunctions of different types of GNRs or chevron-types GNRs have prompted much interest in this field~\cite{joost_correlated_2019, wang_quasiparticle_2012, cai_graphene_2014, rizzo_topological_2018, ma_seamless_2017}. Interestingly, there is not full consensus in the literature regarding the electronic or magnetic properties of simple GNRs, due in large part to the difficulty of describing many-body effects in a sufficiently accurate manner. This paper addresses this issue by considering a GW treatment of the Hubbard model. This method has already been used to describe 2D carbon structures such as GNRs heterojunctions or small polycyclic aromatic hydrocarbons~\cite{joost_correlated_2019, honet_semi-empirical_2021} and it will be applied to finite-size GNRs in this work. 

Armchair graphene nanoribbons (AGNRs) are often categorized into three families, based on their specific electronic band-gap \textit{versus} width dependence~\cite{son_energy_2006, brey_electronic_2006, yang_quasiparticle_2007, lu_competition_2016} and defined by $N_a=3p$, $3p+1$, or $3p+2$ where $N_a$ is the number of atoms across the width of the unit cells and $p$ is an integer. However, tight-binding and $\mathbf {k} \cdot  \mathbf{p}$ approximations predict a zero band-gap (\textit{i.e.}, metallic AGNRs) for the $3p+2$ family~\cite{son_energy_2006, brey_electronic_2006} and the TB approximation leads to the following hierarchy of the gaps: $\Delta_{3p} \ge \Delta_{3p+1} > \Delta_{3p+2}=0$~\cite{son_energy_2006}. In Density Functional Theory (DFT), when considering LDA or the GW correction over LDA, the smallest gaps are also the ones of the $3p+2$ family but they are predicted to be greater than zero (\textit{i.e.}, the AGNRs are non-metallic). What's more, the hierarchy of the two other families is inverted: $\Delta_{3p+1} > \Delta_{3p}$. It is important to note that the LDA + GW method we just invoked implies a GW treatment of the (long-range) Coulombic interaction between electrons. In our study, the GW approximation accounts for the many-body spin up-spin down interactions on the same atomic sites.

There are a several publications addressing the effect of adding a spin-spin interaction term to the TB Hamiltonian~\cite{hancock_generalized_2010, joost_femtosecond_2019, lu_competition_2016}. These studies investigated different parametrizations of an extended TB model (including the Hubbard term) for infinite AGNRs~\cite{hancock_generalized_2010}, the influence of $U$ on the band gap of a finite 7-AGNR~\cite{joost_femtosecond_2019}, and the competition between end states (ES) and bulk states (BS) in different small AGNRs~\cite{lu_competition_2016}. Both Ref.~\onlinecite{lu_competition_2016} and Ref.~\onlinecite{hancock_generalized_2010} employed the MF approximation whereas Ref.~\onlinecite{joost_femtosecond_2019} implemented more advanced self-energy approximations. 

The ES's of finite-size AGNRs are usually understood as topological states, originating from the change in Zak phase $Z_2$ at the interface of the AGNRs and the vacuum~\cite{lawrence_probing_2020,cao_topological_2017,lopez-sancho_topologically_2021}. It has been shown in Ref.~\onlinecite{joost_correlated_2019} that topological states due to such change in Zak phase in GNRs heterostructures are strongly affected by correlation effects. These correlations are thus expected to be of importance in the description of ES in finite-size AGNRs. However, a new topological classification using chiral symmetry was recently proposed and used for GNRs~\cite{jiang_topology_2021,lopez-sancho_topologically_2021}. This classification is based on the $Z$ invariant and predicts the number of topological ES pairs to be $Z$ for semiconducting AGNRs and $Z-1$ for metallic AGNRs.

To the best of our knowledge, Ref.~\onlinecite{lu_competition_2016} is one of the few studies that investigated finite-size AGNRs using a mean-field Hubbard Hamiltonian for different widths and lengths. In this work, we extend the discussion the authors of the study initiated, including correlation \textit{via} the GW approximation as well as comparing the effect of different $U$ values. It appears important to us to study finite-size systems since they ultimately correspond to structures that are experimentally accessible~\cite{wang_giant_2016, lawrence_probing_2020, kimouche_ultra-narrow_2015}. This allows us to compare our theoretical computations to available published experiments. When considering total electronic contributions to the formation of Scanning Tunneling Microscopy (STM) simulations, the GW corrections do not give a clear improvement, in terms of local density of states. However, we show that the spins contributions might be greatly affected when considering the different methods and the present study therefore suggests the experimental verification of the properties of AGNRs via spin-polarized STM in order to access quantities that appear to be most affected by correlation. In contrast, Scanning Tunneling Spectroscopy (STS) experiments yield information that can be readily compared with total density of states and energy positions of the states. We were thus able to compare these quantities for MF and GW with experimental results. The GW corrections allow a much better description of the experimental results. 

The rest of the article is organized as follows. In section~\ref{sec:methods} we introduce the model Hamiltonians as well as the Green's function method used. In section~\ref{sec:GW_U_ES_BS}, we extend the study of Ref.~\onlinecite{lu_competition_2016} on the ES and BS of small AGNRs. We therefore contrast the effect of GW correlation and of  MF approximation considered in Ref.~\onlinecite{lu_competition_2016} as well as the effect of increasing the interaction parameter $U$ of the Hubbard model. In section~\ref{sec:7_GNR_exp}, we compare our results to experiments and DFT computations in terms of local density of states and STM/STS simulations and experiments. We highlight the energy renormalisation of the ES in the GW approximation as well as the stronger localisation of the ES.

\section{Model and methods}
\label{sec:methods}

\subsection{Hubbard model}

The Hubbard model is a popular model used to describe spin interaction effects in materials and in particular in graphene and graphene nanofragments~\cite{castro_neto_electronic_2009, yazyev_emergence_2010, raczkowski_hubbard_2020}. The single-orbital Hubbard Hamiltonian is given by:
\begin{equation}
\hat{H}_{Hubbard} = \Big( -t \sum_{<ij>, \sigma}  \hat{c}^\dagger_{i,\sigma} \hat{c}_{j,\sigma} + hc. \Big) +  U \sum_{i, l} \hat{n}_{i \uparrow} \hat{n} _{i\downarrow},
\label{eq:hubb_ham}
\end{equation}
where indices $i$ and $j$ label atomic sites, and the $\sigma$ index refers to the spin the electron. $\hat{c}^\dagger_{i,\sigma}$ and $\hat{c}_{i,\sigma}$ are the creation and annihilation operators of an electron on site $i$ with spin $\sigma$,  and $\hat{n}_{i \sigma} = \hat{c}^\dagger_{i \sigma} \hat{c}_{i \sigma}$ is the density operator of electron on site $i$ with spin $\sigma$. "$hc.$" stands for "Hermitian conjugate" and $<\ldots>$ indicates that the sum only involves pairs of nearest-neighbour sites. $t$ and $U$ in eq.~(\ref{eq:hubb_ham}) are the only two parameters of the single-orbital Hubbard model. The first one is the hopping parameter while the second one is the interaction parameter. If $U=0$, the Hamiltonian is reduced to the single-orbital tight-binding Hamiltonian~\cite{castro_neto_electronic_2009, yazyev_emergence_2010}.

\subsection{Mean-field approximation}

Treating exactly the Hubbard Hamiltonian for systems of hundreds of electrons is not tractable, due to the fast growing size of the many-body basis set~\cite{honet_exact_2022, jafari_introduction_2008, kingsley_exact_2013}. To circumvent this limitation, a very common approximation of the Hubbard Hamiltonian is the mean-field (MF) approximation where the electron with a given spin on one site interacts with the mean density of the electrons of the opposite spin:
\begin{equation}
\begin{split}
    \hat{H}_{Hub, MF} & = \Big( -t \sum_{<ij>, \sigma}  \hat{c}^\dagger_{i,\sigma} \hat{c}_{j,\sigma} + hc. \Big)  \\
    & +\sum_{ij} U (\hat{n}_{ij\uparrow} \langle \hat{n}_{ij\downarrow} \rangle + \langle \hat{n}_{ij\uparrow} \rangle \hat{n}_{ij\downarrow} ),
\end{split}
\label{eq:Hubbard_MF}
\end{equation}
where $\langle \hat{n}_{il\sigma} \rangle$ is the mean value of the operator $\hat{n}_{il\sigma }$.

The eigen-states of the Hamiltonian in eq.~(\ref{eq:Hubbard_MF}) are found self-consistently, starting from an initial guess for the mean values $\langle \hat{n}_{il\sigma} \rangle$. This approximation has the advantage that it can be expressed in the one-electron basis. It leads to a drastic reduction of the numerical resources needed but presents the caveat of neglecting all the correlation. 

\subsection{GW approximation}

 In order to describe correlated systems beyond the MF but at a smaller computational cost than that of the exact diagonalization, a number of approximations from the non-equilibrium Green's function framework have been developed such as T-matrix approximation, second-order Born approximation, GW approximation, \textit{etc}.~\cite{schlunzen_nonequilibrium_2016, reining_gw_2018, di_sabatino_reduced_2015, romaniello_self-energy_2009, romaniello_beyond_2012}. In this study, we adopt the GW approximation, whose foundation is Dyson's equation:

\begin{equation}
    G^{R} (\omega) = G^{R}_0 (\omega) + G^{R}_0 (\omega) \Sigma^{R} (\omega) G^{R} (\omega),
\label{Dyson_equ}
\end{equation}
where $G^{R}_0$ is the non-interacting retarded Green's function (computed using the MF solution), $G^{R}$ is the exact retarded Green's function, and $\Sigma^{R} $ the retarded self-energy.

In the single-orbital basis, Dyson's equation is written in matrix form such that all quantities are matrices and products are matrix products. It is a frequency (\textit{i.e.}, energy) dependent equation. Note that we work in natural units, such that $\hbar = 1$ and $\omega$ is in energy units. The core of the GW description consists in approximating the self-energy as the product of $G$ (Green's function) and $W$ (screened potential). Dyson's equation is then solved self-consistently by computing an updated Green's function at each iteration, starting from the MF one. More details on the method and its implementation can be found in Refs.~\onlinecite{honet_semi-empirical_2021,honet_exact_2022, joost_correlated_2019}.

\subsection{Observables from Green's functions}

One can obtain spectral properties from either Green's function ($G_0$ or $G$). This includes local and total density of states ($LDOS$ or $n_{i\sigma}(\omega)$ and $DOS$ or $D(\omega)$). These quantities are expressed in terms of the Green's function as:

\begin{equation}
n_{i\sigma}(\omega) = \frac{1}{2\pi} A_{i\sigma,i\sigma} (\omega)
\label{eq:ldos_GF}
\end{equation}
and
\begin{equation}
D(\omega) = \sum_{i\sigma} n_{i\sigma}(\omega),
\label{eq:dos_GF}
\end{equation}
where $A_{i\sigma,j\sigma'} (\omega) = -2 \Im(G^R_{i\sigma,j\sigma'}(\omega))$ is the spectral function.

The local-density of states can be accessed experimentally as spatially-resolved $dI/dV$ images using STS. Working in interval of energies $[E_1,E_2]$, the calculation of $dI/dV$ involves non-diagonal terms of the Green's function and is expressed as~\cite{joost_correlated_2019, meunier_tight-binding_1998}:
\begin{equation}
\frac{\dd{I}}{\dd{V}} (x,y,z_0) = \int_{E_1}^{E_2} \dd{\omega} \sum_{ij} A_{ij}(\omega) z_0^2 e^{\lambda^{-1} \abs{\vec{r}-\vec{r_i}}} e^{\lambda^{-1} \abs{\vec{r}-\vec{r_j}}},
\label{eq:stm_simu}
\end{equation}
where $z_0$ is the tip's height of the simulated $STS$; $\lambda$ is a length parameter that accounts for the spatial extension of localized orbitals; and $(x,y,z_0)=\vec{r}$ is the location where the $STS$ is simulated and $\vec{r}_i$ are the atomic positions. We used the values $z_0 = 0.4 \hspace{0.1cm} d_{CC}$ and $ \lambda=1.72  \hspace{0.1cm} d_{CC}$ (with $d_{CC} = 0.142 \hspace{0.1cm} \rm nm$ the carbon-carbon distance in graphene) throughout this paper. The influence of these parameters on the simulated STS is a bit discussed in th SI.

\section{Effect of GW and $U$ on the competition between topological end states and bulk states}
\label{sec:GW_U_ES_BS}

This section is devoted to the theoretical investigation of the effect of using the GW approximation and of changing the Hubbard interaction parameter $U$ on the properties of one representative AGNR system of each family. To facilitate a comparison with a previous study, we focus on 7-AGNR, 9-AGNR, and 11-AGNR since they have been studied in the TB and MF approximations of the Hubbard model in Ref.~\onlinecite{lu_competition_2016}. The authors of this research listed a number of observations that we will further discuss here. For instance, it was observed that:
\begin{itemize}
    \item In contract to the ES, the BS energies are almost unaffected by the interaction term of the Hubbard model ($U$) when compared to the TB energies.
    \item The Highest Occupied Molecular Orbitals (HOMOs) and Lowest Unoccupied Molecular Orbitals (LUMOs) are ES for AGNRs with ${\rm mod}(n,3)=1$ and ${\rm mod}(n,3)=0$, while they can be BS for AGNRs with ${\rm mod}(n,3)=2$. In particular, the HOMOs and LUMOs are ES for 7-AGNRs and 9-AGNRs, with the difference between BS and ES in 9-AGNRs being smaller than in 7-AGNRs. In contrast, the HOMOs and LUMOs can be ES or BS (if the ribbon is long enough) for 11-AGNRs.
\end{itemize}

The number of ES can be predicted in the TB approximation in terms of the chiral topological invariant $Z$~\cite{lopez-sancho_topologically_2021}. Both 7- and 9- AGNRs have $Z=1$ and, since they are semiconducting, they are predicted to host $Z=1$ pair of topological ES. We note that 11-AGNRs have $Z=2$ but because they are metallic, they are predicted to have $Z-1=1$ pair of topological ES. All studied AGNRs in this section are thus predicted to have one pair of topological ES at the TB level.

\subsection{Effects of GW correction with $U=4/3t$}
\label{sec:GW_effect}

\begin{figure}
\centering
    \includegraphics[width=8.5cm]{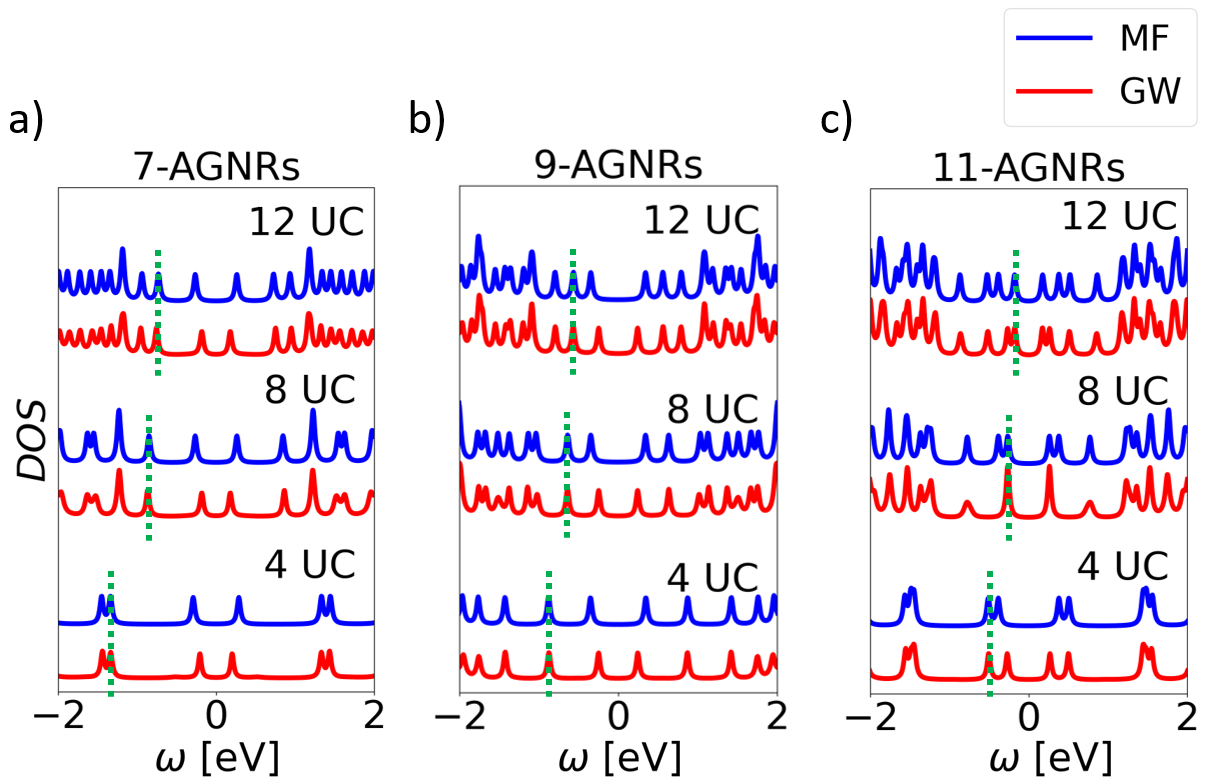}
\caption{Density of states for 7-AGNRs (a), 9-AGNRs (b), and 11-AGNRs (c). All panels display DOS for three lengths: 4 UC (two bottom curves), 8 UC (two middle curves) and 12 UC (two top curves). The DOS are given for the MF (blue) and GW (red) approximations of the Hubbard model with $t=2.7~eV$ and $U=4/3~t$. The highest occupied bulk states are indicated with the green dotted lines, the other frontier states (below $E_F$) peaks corresponding to ES. All Fermi levels have been shifted to $0~eV$ for better visualisation.}
\label{fig:DOS_7_9_11_U_1p33t}
\end{figure}

First, we start by considering the systems studied in Ref.~\onlinecite{lu_competition_2016} with the same model parameters, \textit{i.e.} $t=2.7~eV$ and $U=3.6~eV = 4/3~t$, in order to assess the influence of the GW correction. Fig.~\ref{fig:DOS_7_9_11_U_1p33t} shows the DOS obtained in the MF and GW approximations for 7-AGNRs (fig.~\ref{fig:DOS_7_9_11_U_1p33t} (a)), 9-AGNRs (fig.~\ref{fig:DOS_7_9_11_U_1p33t} (b)), and 11-AGNRs (fig.~\ref{fig:DOS_7_9_11_U_1p33t} (c)) for three different lengths: 4, 8, and 12 unit cells (UC). 

The HOMO and LUMO states of 7-AGNRs and 9-AGNRs are ES in both the MF approximation and GW correction (see fig.~\ref{fig:DOS_7_9_11_U_1p33t}, where the green dotted lines mark the BS and the other frontier states corresponding to ES). The effect of GW compared to MF for all AGNRs considered is mainly to shift the ES towards $E_F$, leaving the energies of BS almost unchanged. For the 11-AGNRs however, it was already observed in Ref.~\onlinecite{lu_competition_2016} that the HOMO and LUMO could be BS and not ES: this is the case for the 8 and 12 UC 11-AGNRs in the MF approximation (see fig.~\ref{fig:DOS_7_9_11_U_1p33t} and Ref.~\onlinecite{lu_competition_2016}). A crossing between ES and BS thus occurs between the 4 UC and the 8 UC systems. In GW, since the BS energies are almost unchanged and the ES are shifted towards $E_F$, the crossing occurs for larger system length: while the crossing already occurred for the 8 UC in MF, the GW energies of ES and BS are very close to each other for this same length.

\subsection{Effect of increasing the $U$ parameter}

In previous GW studies on carbon nanostructures~\cite{honet_semi-empirical_2021, joost_correlated_2019}, the interaction parameter $U$ had to be taken larger than the value used in Ref.~\onlinecite{lu_competition_2016} and in section~\ref{sec:GW_effect} to match available experimental data. This is also the case here, as we will show below (see sec.~\ref{sec:7_GNR_exp}). We thus now turn our interest to the effect of increasing this parameter from $U=4/3~t $ to $U=2~t$.

\begin{figure}
\centering
    \includegraphics[width=8.5cm]{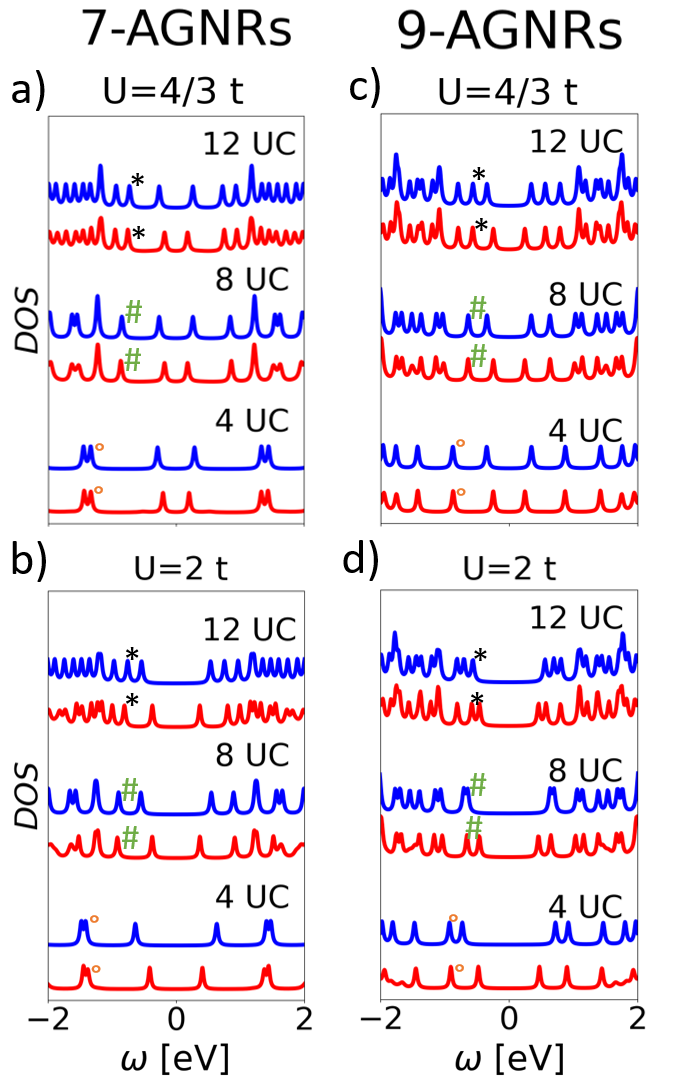}
\caption{Density of states for 7-AGNRs (left) and 9-AGNRs (right) of three lengths: 4 UC, 8 UC, and 12 UC. The DOS are given for MF (blue) and GW (red) approximations of the Hubbard model with $t=2.7~eV$ and $U=4/3~t$ for the upper panels and $U=2~t$ for the lower panels. The black * (resp. green $\#$ and orange °) indicate BS of the 12 UC (resp., 8 UC and 4 UC) that are almost unchanged in energy. All Fermi levels have been shifted to $0~eV$ for better visualisation.}
\label{fig:DOS_7_9_U_1p33t_2t}
\end{figure}

Figs.~\ref{fig:DOS_7_9_U_1p33t_2t} a) and b) show the DOS for the three selected lengths of 7-AGNRs in the MF and GW approximations. We observe that the BS energies are very similar for the two values (see peaks indicated in the figure). As for $U=4/3~t $, the ES's are more affected and shifted away from $E_F$ when increasing the $U$ parameter. This shift is higher in MF than in GW. In the particular case of 7-AGNRs, all HOMOs and LUMOs remain ES as in the case of $U=4/3~t$.

Turning to 9-AGNRs, the same global trends are observed as for 7-AGNRs (see fig.~\ref{fig:DOS_7_9_U_1p33t_2t} c) and d)). However, since the initial energy spacing between ES and BS is smaller for 9-AGNRs, a crossing between ES and BS is observed in the MF approximation from the 4 UC system to the 8 UC system. It appears that HOMOs and LUMOs are now BS for the 8 UC and 12 UC 9-AGNRs in the MF. The GW approximation however, still predicts the HOMOs and LUMOs to be ES for these systems. We thus show that inversions between BS and ES may occur in 9-AGNRs in addition to the observed inversion in Ref.~\onlinecite{lu_competition_2016} for 11-AGNRs.

The case of 11-AGNRs shows several similarities with the two first systems (see fig.~1 in the SI for the densities of states). The main effect of increasing $U$ is to shift the ES away from $E_F$. In the 11-AGNRs, the shift for the BS in MF is larger than for other systems but is still small compared to the shift of ES, being always more than three times smaller. This shift for BS is reduced when considering the GW approximation. The consequence of these observations is that the crossing between BS and ES occurs for smaller systems when increasing $U$, as for the 9-AGNRs. In MF, none of the three considered lengths exhibits ES as HOMO and LUMO, whereas it is only the case for the 4 UC system in GW (HOM0s and LUMOs being BS for 8 UC and 12 UC in GW).

\section{7-AGNRs properties}
\label{sec:7_GNR_exp}

We now focus on the 7-AGNRs that have been recently synthesized, characterized by STM/STS and studied theoretically by the means of \textit{ab initio} simulations \cite{wang_giant_2016}. More specifically, the energy splitting of end-states $\Delta_{ZZ}$ in these nanoribbons was shown to be of significant magnitude compared with the bulk-states gap $\Delta_{AC}$. More precisely, the values $\Delta_{ZZ} \simeq 1.9-2 eV$ and $\Delta_{AC} \simeq 3-3.5 eV$ were found for the different investigated sizes of 7-AGNR (see Fig. 3 of Ref.~\onlinecite{wang_giant_2016}). This reference proposed an experimental strategy to transfer the nanoribbons from a Au(111) on which they were synthesized onto a NaCl monolayer, which is itself deposited on a a Au(111) substrate. This experimental proposal to make GNRs neutral and electronically decoupled from Au(111) is expected to approach free-standing properties of GNRs~\cite{kharche_width_2016}, this is why we considered this experimental study in particular.

\begin{figure*}
\centering
    \includegraphics[width=17cm]{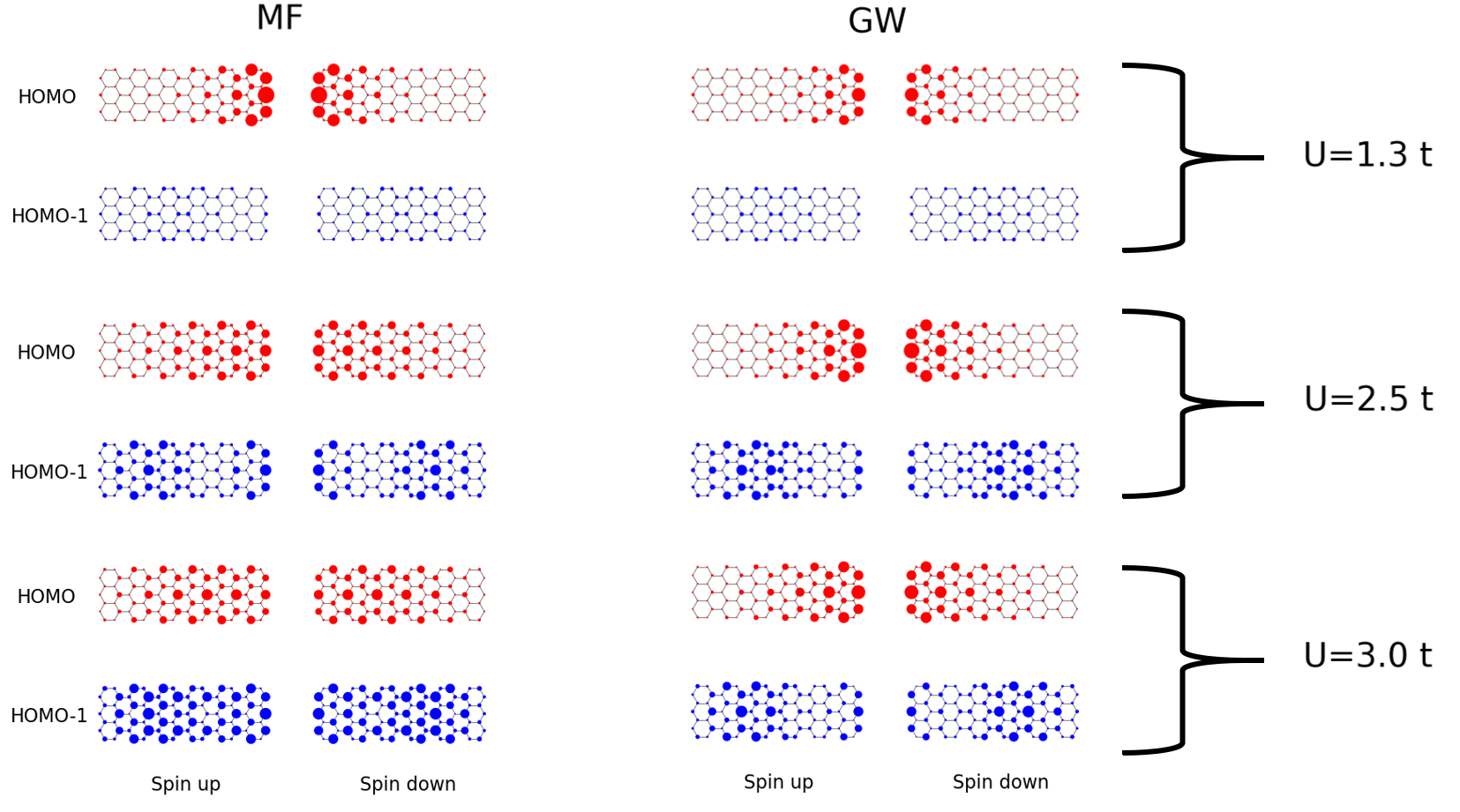}
\caption{LDOS of HOMO (red dots) and HOMO-1 (blue dots) of a 6 unit cells 7-AGNR in the MF (left) and GW (right) approximations for three values of the $U$ parameter: $1.3t$, $2.5t$, and $3t$ from top to bottom. In each approximation, the spin-up and spin-down electron densities are represented separately, at the left and the right respectively. The red dots correspond to HOMO and the blue ones to HOMO-1. The sizes of the dots are proportional to the mean occupation of the atomic sites.}
\label{fig:LDOS_7_GNR_6_UC}
\end{figure*}

The local densities of states (LDOS) for the nearest peak below the Fermi level -- which has been identified as being HOMO -- and the peak of energy right under the HOMO (HOMO-1) are given at fig.~\ref{fig:LDOS_7_GNR_6_UC} for the MF and GW approximations with $U$ parameters of $1.3t$, $2.5t$, and $3t$. For $U$ parameters of the order of magnitude of $t$ ($U=1.3t$ on fig.~\ref{fig:LDOS_7_GNR_6_UC}), the LDOS are hardly changed from MF to GW. Both MF and GW are in great agreement with DFT results from Ref.~\onlinecite{wang_giant_2016} and the HOMO is predicted to be an edge states while the HOMO-1 is predicted to be a bulk state (spread over the ribbon). When increasing $U$ in MF, the HOMO tends to be less end-localized and, for $U=3t$, the atoms at the middle of the two ends of the ribbons are not the most occupied atomic sites by the HOMO anymore. On the contrary, the HOMO-1 in MF tends to be more localized at the edges for larger values of $U$. This state also becomes less spatially symmetric for each spin channel. One observed that the effect of GW corrections is to further keep the edge-localized character of the HOMO state while $U$ is increasing. This characteristic has the consequence that a larger range of $U$ values could be considered when describing 7-AGNRs with the constraint of the HOMO and LUMO states to be edge-localized, as observed experimentally~\cite{wang_giant_2016}. The extension of the possible $U$ values to match with experiment or other methods was previously highlighted by us in Ref.~\onlinecite{honet_exact_2022} where we showed that the outset of an artificial phase transition induced by MF was shifted towards higher $U$ values in GW, in a better agreement with exact results. We notice that for high $U$ values, none of the approximations really reproduces the DFT results of Ref.~\onlinecite{wang_giant_2016}. However, DFT is also known for not taking into account correlation effects correctly.

\begin{figure*}
\centering
    \includegraphics[width=17cm]{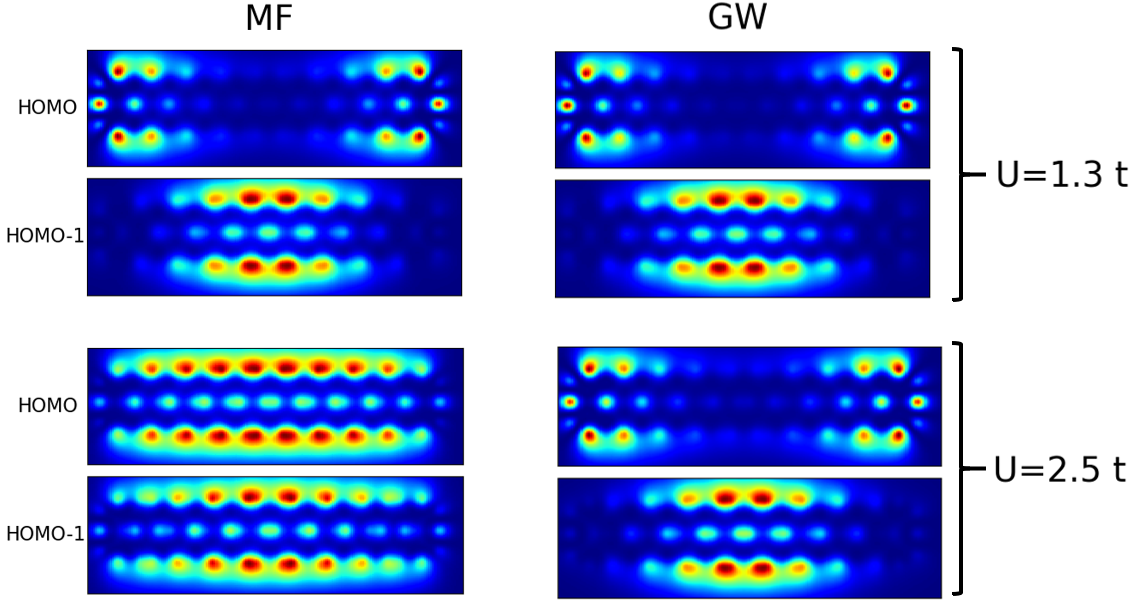}
\caption{STM simulations of HOMO and HOMO-1 of a 10 unit cells 7-AGNR in the MF (left) and GW (right) approximations for $U=$ $1.3t$ (top) and $2.5t$ (bottom). HOMO STM simulations are computed in the intervals of energy $[-0.099t, -0.085t]$ for MF and $[-0.071t, -0.058t]$ for GW while the intervals of energy for the HOMO-1 simulations are $[-0.292t, -0.278t]$ for MF and $[-0.301t, -0.288t]$ for GW.}
\label{fig:STM_7_GNR_6_UC}
\end{figure*}

The simulated $STM$ images of a longer 7-AGNR (10 unit cells) are shown in figure~\ref{fig:STM_7_GNR_6_UC} for $U$ values of $1.3t$ and $2.5t$. Since $STM$ simulations take into account both spin channels, we see that despite the fact that the spatial asymmetry of the LDOS increases for each spin, the total simulated STM maps remain essentially identical in GW between $U=1.3t$ and $U=2.5t$, while they are strongly modified in MF. Only the STM simulations corresponding to the edge-located HOMO are in agreement with the experimental data reported in Ref.~\onlinecite{wang_giant_2016}, that is the one for small $U$ ($1.3t$) in MF and both simulated STM maps for GW. Since the spatial asymmetry increases in spin channel for the LDOS (see Fig. \ref{fig:LDOS_7_GNR_6_UC}), we believe that spin-polarized STM experiments could help determine the amount of correlation and compare different level of theory (model Hamiltonians and \textit{ab initio}).

\begin{figure*}
\centering
    \includegraphics[width=17cm]{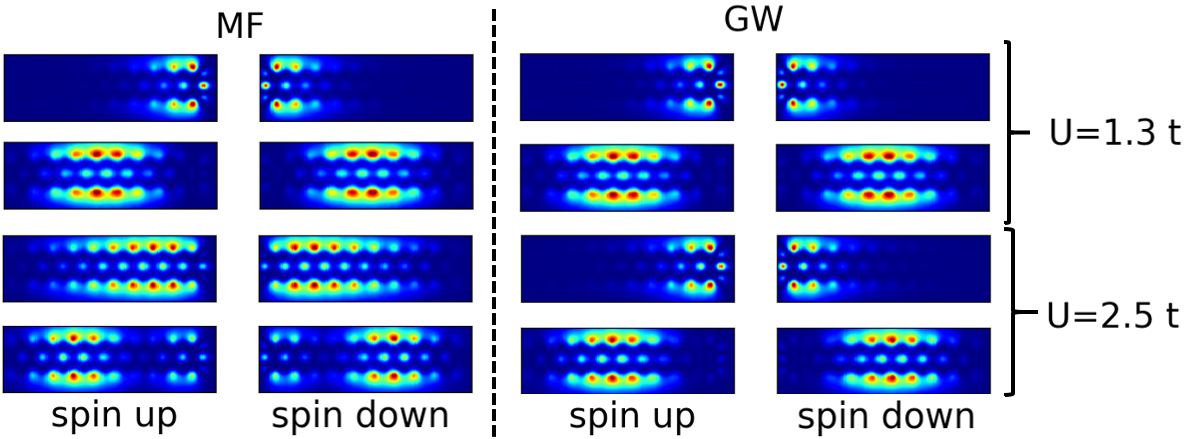}
\caption{Spin-resolved STM simulations of HOMO and HOMO-1 of a 10 unit cells 7-AGNR in the MF (left) and GW (right) approximations for $U=$ $1.3t$ (top) and $2.5t$ (bottom). In each approximations, the spin-up (spin-down) channel are the left (right) images. The intervals in energy are the same as for fig.~\ref{fig:STM_7_GNR_6_UC}.}
\label{fig:STM_7_GNR_6_UC_spins}
\end{figure*}

The experimental spatial STM maps can be well reproduced by both MF and GW  (figure~\ref{fig:STM_7_GNR_6_UC}) and one  cannot determine which is a better approximation from these simulations. However, when looking at STM simulations of separate spin channels, a clearer difference can be observed for results obtained with small $U$ values ($U=1.3t$ on fig.~\ref{fig:STM_7_GNR_6_UC_spins}) and larger $U$ values in the GW approximation ($U=2.5t$ on fig.~\ref{fig:STM_7_GNR_6_UC_spins}). This should be understood as a fingerprint of the LDOS studied before and showed in fig.~\ref{fig:LDOS_7_GNR_6_UC}. We therefore suggest that further experimental exploration of GNRs including spin-polarized STM experiments could be used, with the support of theoretical simulations, to look for the expression of many-body correlation effects. Moreover, it appears that a change in the parameters of the simulations such as the local orbital extension (or the height of the tip) could reveal more accurately the LDOS features (see figs. 2 and 3 in the SI). These parameters can be used to describe different experimental set-ups such that there should exist experimental STM measurement parameters approaching LDOS features.

\begin{figure}
\centering
    \includegraphics[width=8.5cm]{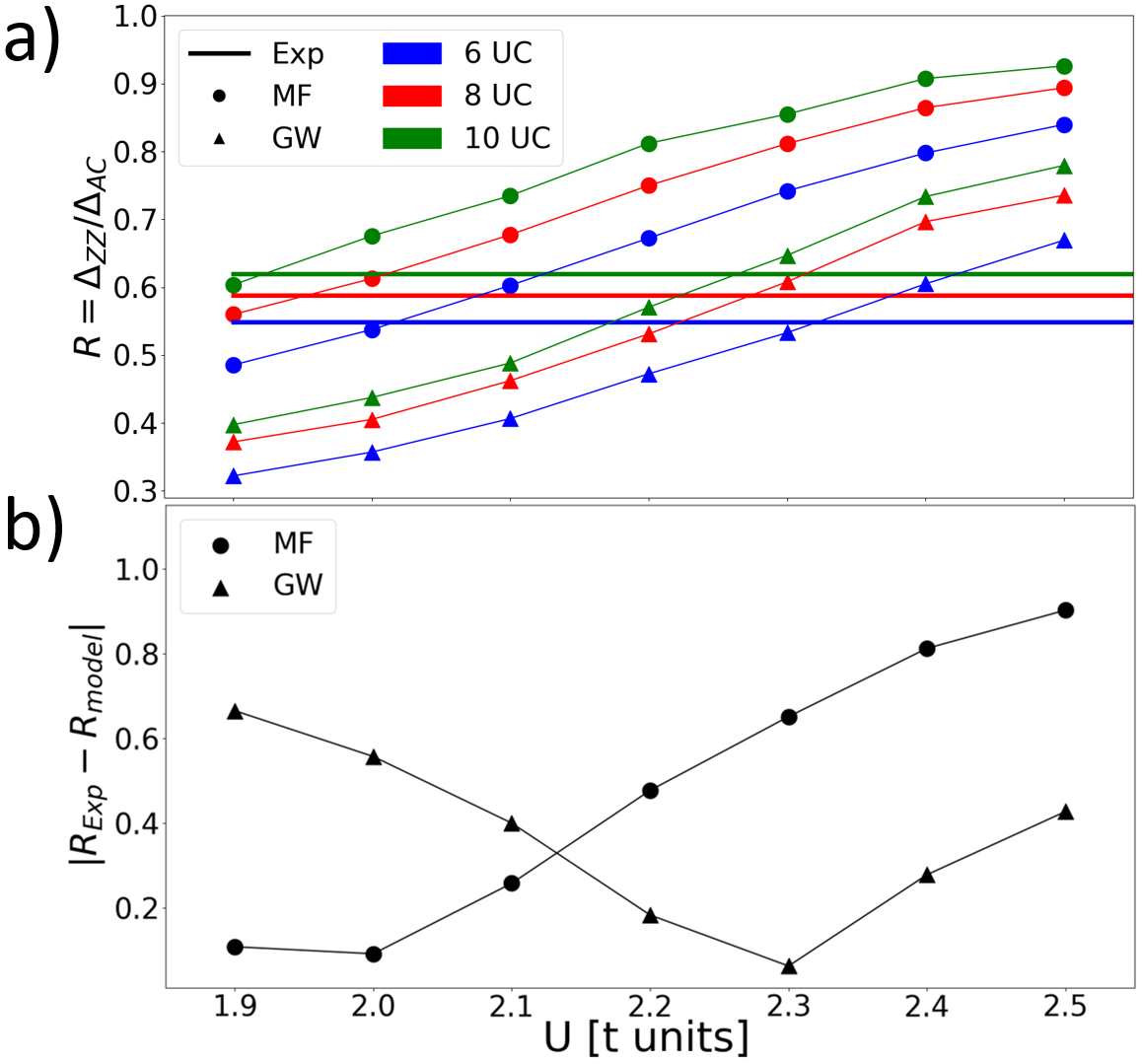}
\caption{a) MF (circles) and GW (triangles) values of the ratio $R=\Delta_{ZZ}/\Delta_{AC}$ as a function of the $U$ parameter for 6 units cells (blue), 8 unit cells (red), and 10 unit cells (green). The constant horizontal lines represent the values of the experiment, extracted from Ref.~\onlinecite{wang_giant_2016} with WebPlotDigitizer. b) Absolute value of the errors in the $R$ ratio for MF (circles) and GW (triangles), compared to experimental results, as a function of $U$.}
\label{fig:Delta_ZZ_Delta_AC}
\end{figure}

We now focus on the simulation of STS measurements to describe the edge-states gap ($\Delta_{ZZ}$) as well as the bulk-states gap ($\Delta_{AC}$). In the Hubbard model, for a given $U/t$ ratio, the energy range of all the spectra can be scaled via the $t$ parameter. Instead of focusing on absolute values of the two gaps ($\Delta_{ZZ}$ and $\Delta_{AC}$), we consider the ratio $R=\Delta_{ZZ}/\Delta_{AC}$, comparing it to experimental values of Ref.~\onlinecite{wang_giant_2016} (see figure 3.g. therein), reproduced as horizontal lines on fig.~\ref{fig:Delta_ZZ_Delta_AC}~a). The evolution of $R$ for the MF and GW approximations as a function of $U$, compared with the experimental ones for 7-AGNRs with $6, 8$ and $10$ unit cells is shown in figure~\ref{fig:Delta_ZZ_Delta_AC}~a). We see in this figure that for both approximations and for all lengths, $R$ increases with $U$. The best $U$ values are $U=2t$ and $U=2.3t$ for MF and GW approximations respectively.

For an easier global comparison, the sums (over the three different lengths) of the absolute values of the errors of the $R$ ratio of both approximations are compared with experiment in fig.~\ref{fig:Delta_ZZ_Delta_AC}~b). This confirms that the best $U$ values are $U=2t$ for MF and $U=2.3t$ for GW. Even if the agreement is slightly better at the GW minimum than at the MF minimum, the improvement does not appear to be significant, even if one can already see in fig.~\ref{fig:Delta_ZZ_Delta_AC}~a) that the experimental ratios are reproduced better by GW with $U=2.3t$ than by MF with $U=2t$. An additional benefit of this comparison is to fix the $U$ values at $U=2t$ for MF and $U=2.3t$ for GW for the following of this comparison with the experiment.

\begin{figure*}
\centering
    \includegraphics[width=17cm]{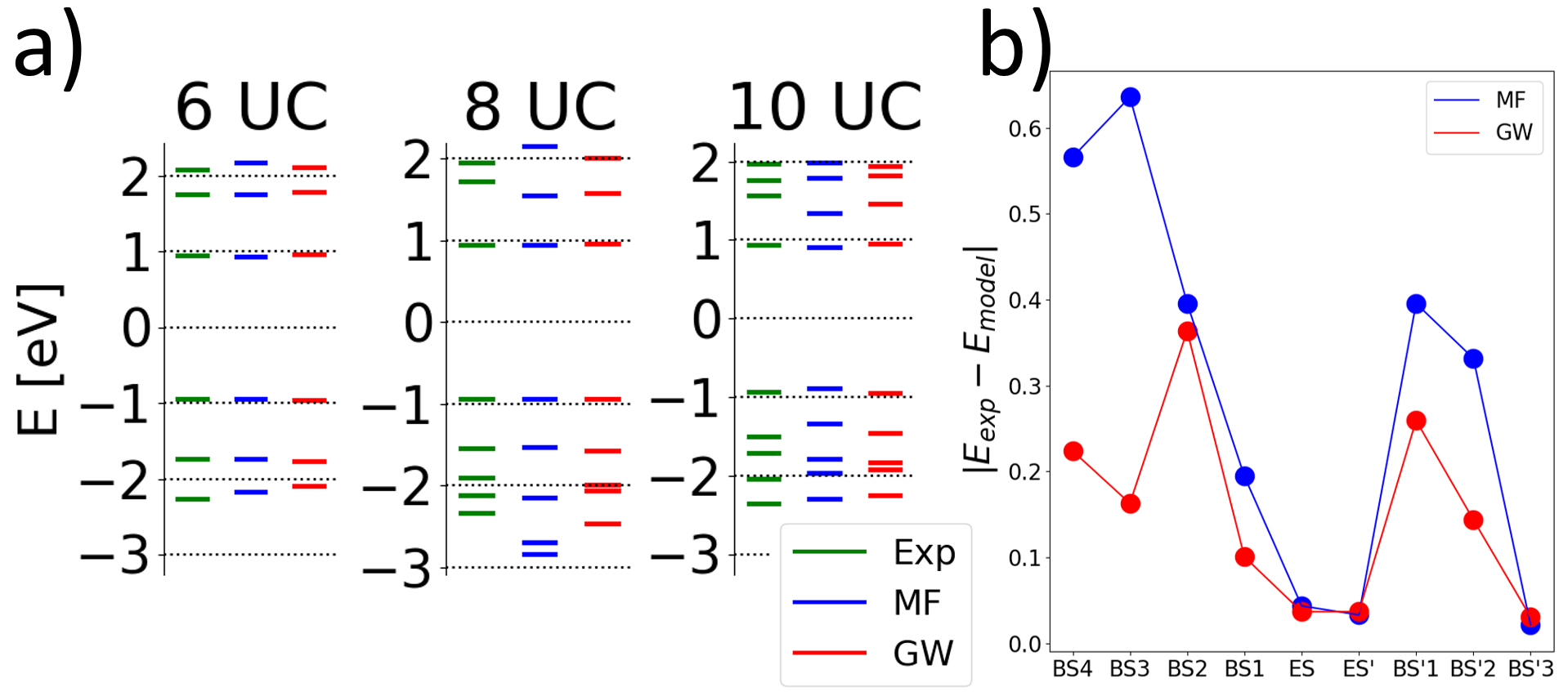}
\caption{a) Energy diagrams for 7-AGNRs of three different lengths (6, 8 and 10 UC). In green, experimental data of Ref.\onlinecite{wang_giant_2016}, extracted with WebPlotDigitizer. In blue, MF results for the parameters $t=4.46 eV$ and $U=2t$ and in red, GW for the parameters $t=4.22 eV$ and $U=2.3t$. b) Sum of errors between MF (blue) or GW (red) and experimental data. The states are labelled with a " ' " symbol for unoccupied states and with a number for BS, increasing with the separation with the Fermi level (located at the middle of the two ES. The sum is carried over the three lengths including all accessible experimental data, but not all the lengths are included for all peaks. }
\label{fig:All_orbitals}
\end{figure*}

Having fixed these parameter ratios, we extended our comparison between MF and GW on one hand and experiment on the other, to include more peaks beyond the two gap values $\Delta_{ZZ}$ and $\Delta_{AC}$. Fig.~\ref{fig:All_orbitals}~a) shows the energy diagrams for the three different lengths of 7-AGNRs for the experiment, the MF and GW approximations for all peaks accessible from fig.~3 of Ref~\onlinecite{wang_giant_2016} (6, 8, and 9 peaks for the 6 UC, 8 UC and 10 UC systems respectively). Fig.~\ref{fig:All_orbitals}~b) shows the errors between MF or GW approximations and experiments. The errors have been summed up for 6, 8 and, 10 unit cells 7-AGNRs in the limits of available experimental data (some lower and higher energy peaks are not visible for shorter 7-AGNRs in Ref.~\onlinecite{wang_giant_2016}). The hopping parameters $t=4.46 eV$ for MF and $t=4.22 eV$ have been found to minimize the total errors, so that we decided to show the results for these values. From the $R$ ratios shown in fig.~\ref{fig:Delta_ZZ_Delta_AC}, we concluded that the agreement for GW was slightly better than for MF. It is clear now that the inclusion of correlations in GW leads to better energy results than the MF approximation.

We note that the hopping terms to reproduce the experimental results in both approximations are relatively large (between $4 eV$ and $4.5 eV$) compared to values usually found in the literature to describe graphene (between $2.7 eV$ and $3.2 eV$)~\cite{castro_neto_electronic_2009, yazyev_emergence_2010, bullard_improved_2015}. This is related to the renormalization of the hopping integral in a Hubbard model. Indeed, the typical way to find a value for the hopping parameter is to compute electrical properties such as the dispersion relation and to compare and fit it to \textit{ab initio} computations~\cite{castro_neto_electronic_2009, schuler_2016}. In these works, the dispersion relation are compared between the tight-binding approximation and~\textit{ab initio} computations, meaning that the interaction term of the Hubbard Hamiltonian is not included. The inclusion of $U$ modifies the relation dispersion of graphene and can open a band gap~\cite{raczkowski_hubbard_2020}.

The hopping term in the Hubbard Hamiltonian is responsible for electron delocalization. On the other hand, local electronic interaction introduced with the $U$ parameter tends to induce a localisation of the electrons~\cite{in_t_Veld_2019}. Adding non-local interaction terms (limited to nearest-neighbour), one ends up with the extended Hubbard model (EHM) defined by:

\begin{equation}
    H_{EHM} = \big( -t \Sigma_{\langle i,j \rangle, \sigma} c^\dagger_{i\sigma} c_{j\sigma} + hc. \big) + U \Sigma_i n_{i\uparrow} n_{i\downarrow} + \frac{V}{2} \Sigma_{\langle i,j \rangle} n_i n_j,
    \label{eq:Ham_EHM}
\end{equation}
with $V$ being the non-local nearest-neighbour interaction parameter.

The non-local interaction term tends to delocalize the electrons as the hopping term and it has been shown that the EHM can be mapped towards an effective Hubbard model, setting $V=0$ and decreasing the ratio $U/t$~\cite{in_t_Veld_2019, schuler_optimal_2013}. This can be understood as switching off a term that tends to delocalize electrons (non-local interactions) but increasing the other term responsible for delocalization (the hopping term). The EHM was studied for graphene in Ref.~\onlinecite{schuler_optimal_2013} and the following parameters were found: $t=2.8 eV$, $U=3.63 t$, and $V=2.03 t$. Using these values in the proposed mapping of Ref.~\onlinecite{in_t_Veld_2019}, one ends up with an effective Hubbard model with parameters $t=3.16 eV$ to $t=5.80eV$ and $U=1.82 t$ to $U=3.33 t$. These values are compatible with the ones we used in this paper.

To sum up: the seemingly large hopping parameter found to optimize the match with experiment can be explained by the fact that we neglected non-local interactions and used the Hubbard model (introducing local interactions) and, doing so, the hopping parameter has to be renormalized, according to Refs.~\onlinecite{in_t_Veld_2019} and~\onlinecite{schuler_optimal_2013}.

\section{Conclusion}

In summary, we applied the GW approximation developed in Refs.~\onlinecite{honet_semi-empirical_2021, joost_correlated_2019} to study the effect of a many-body Hubbard-type treatment of finite-size AGNRs. We first extended the theoretical study of Ref.~\onlinecite{lu_competition_2016} that considered the description of the AGNRs using the Hubbard model in the MF approximation for an interaction parameter $U=4/3t$. We showed that the GW approximation mainly affects the ES whereas the BS are less affected. We observed that GW tends to reduce the gap between end states in the considered AGNRs, affecting the effect of "competition" between ES and BS to be the HOMO and LUMO. In particular, the crossing between these states when increasing the length of a 11-AGNR turned out to be at smaller length in GW than in MF. We then explored the effect of increasing the $U$ interaction term of the Hubbard model. The main effect was also observed on the ES, increasing the ES gap ($\Delta_{ZZ}$) and therefore changing the ratio between ES and BS gaps ( $\Delta_{ZZ} /\Delta_{AC} $) as well as potentially affecting the HOMO and LUMO. More specifically, we showed that it is possible to have BS as HOMO and LUMO in 9-AGNRs, in the MF approximation. It is not the case for GW, that predicts a larger range of $U$ resulting in ES as HOMO and LUMO.

We considered in more details the 7-AGNRs and compared our model predictions to experimental data of recently synthesized 7-AGNRs of Ref.~\onlinecite{wang_giant_2016}. This methodology yields very different results than other experimental studies that we were able to quantitatively reproduce theoretically within the GW approximation with very good agreement. The parameters found in our study are significantly larger than the ones usually employed in TB or Hubbard model for the description of graphene~\cite{castro_neto_electronic_2009, yazyev_emergence_2010, bullard_improved_2015, joost_correlated_2019}. Whereas DFT based methods are largely used to describe carbon nanosystems, they are known to not account accurately for correlation effects, specifically the spin-spin correlation considered here in the Hubbard model. 

\section*{Acknowledgements}

A.H. is a Research Fellow of the Fonds de la Recherche Scientifique - FNRS. This research used resources of the "Plateforme Technologique de Calcul Intensif (PTCI)" (\url{http://www.ptci.unamur.be}) located at the University of Namur, Belgium, and of the Université catholique de Louvain (CISM/UCL) which are supported by the F.R.S.-FNRS under the convention No. 2.5020.11. The PTCI and CISM are member of the "Consortium des Équipements de Calcul Intensif (CÉCI)" (\url{http://www.ceci-hpc.be}).


\bibliographystyle{ieeetr}
\bibliography{bibliography}

\end{document}